\documentclass[stef,twoside]{stefano}

\usepackage{amsmath}
\usepackage{amssymb}
\usepackage{graphics}
\usepackage{rotating}
\usepackage{cite}
\usepackage{color}

\def\makeheadbox{{%
\hbox to0pt{\vbox{\baselineskip=10dd\hrule\hbox
to\hsize{\vrule\kern3pt\vbox{\kern3pt \hbox{{\sc Eur. Phys. J. C} {\bf 63},
157-162 (2009)} \hbox{ {\sc
{\color{blue}{dma}}[{\color{black}{imecc}}]{\color{red}{UniCamp}} }
\hspace*{10.3cm} {\color{blue}{$\boldsymbol{\Sigma \delta \Lambda}$ }}}
\kern3pt}\hfil\kern3pt\vrule}\hrule}%
\hss}}}

\def\0{\mbox{\tiny $0$}}
\def\1{\mbox{\tiny $1$}}
\def\2{\mbox{\tiny $2$}}
\def\3{\mbox{\tiny $3$}}
\def\4{\mbox{\tiny $4$}}
\def\5{\mbox{\tiny $5$}}
\def\6{\mbox{\tiny $6$}}
\def\7{\mbox{\tiny $7$}}
\def\8{\mbox{\tiny $8$}}
\def\9{\mbox{\tiny $9$}}
\def\min1{\mbox{\tiny $(-\,1)$}}
\def\m2{\mbox{\tiny $(-\,2)$}}
\def\={\mbox{\tiny $=$}}
\def\T{\mbox{\tiny T}}
\def\R{\mbox{\tiny R}}
\def\mi{\mbox{\tiny $-$}}
\def\pl{\mbox{\tiny $+$}}

\begin{document}
%

\title{\Large PLANAR DIRAC DIFFUSION}

\author{
Stefano De Leo \inst{1}
\and Pietro Rotelli\inst{2} }

\institute{
Department of Applied Mathematics, University of Campinas\\
PO Box 6065, SP 13083-970, Campinas, Brazil\\
{\em deleo@ime.unicamp.br}\\
 \and
Department of Physics, University of
Lecce and INFN Lecce\\
PO BOX 193, CAP 73100, Lecce, Italy\\
{\em rotelli@le.infn.it} }


\date{Submitted {\em April, 2009}. Revised and Accepted {\em July 2009}.}

\abstract{We present the results of the planar diffusion of a
Dirac particle by step and barrier potentials, when the incoming
wave impinges at an arbitrary angle with the potential. Except for
right-angle incidence this process is characterized by the
appearance of spin flip terms. For the step potential, spin flip
occurs for both transmitted and reflected waves. However, we find
no spin flip in the transmitted barrier result. This is surprising
because the barrier result may be derived directly from a two-step
calculation. We demonstrate that the spin flip cancellation indeed
occurs for each ``particle" (wave packet) contribution.}


\PACS{ {03.65.Pm} {({\sc pacs}).}}













\titlerunning{\sc planar dirc diffusion}

\maketitle


\section*{I. INTRODUCTION}

In a previous paper\cite{Ddif}, we investigated the
one-dimensional phenomena of diffusion of a Dirac particle from
step and barrier potentials. One of the first observations made in
that work was the simplifying fact that spin flip did not occur
for either potential. This result was not limited to the non
relativistic limit where it might have been expected. It is
however, as we shall show below an exceptional result.

We consider in this paper potentials and particle rays situated in
a plane, the $y$-$z$ plane. The potentials are functions of only
one variable, i.e. $V=V(z)$,  while the incoming particle
direction lies in the $y$-$z$ plane with $\theta$  the impact
angle with the potential. The outgoing wave momenta, be they
reflected, transmitted or those in the barrier region, must of
course also lie in the $y$-$z$ plane.

\begin{center}
\begin{picture}(300,150) \thinlines
\put(50,45){\line(0,1){30}} \put(48,79){$p_{\2}$}
\put(50,75){\line(1,0){30}} \put(50,45){\vector(1,1){30}}
\put(82,79){$\boldsymbol{p}$} \put(50,45){\line(1,0){30}}
\put(83,43){$p_{\3}$} \put(60,46.5){$\theta$}
 \put(80,45){\line(0,1){30}}
\put(212,10){\vector(0,1){58}} \put(210,76){$y$}
\put(212,10){\vector(1,1){38}} \put(252,52){$x$}
\put(62,10){\vector(1,0){210}} \put(280,8){$z$}
\put(50,120){\mbox{ $V(z)\,=\,\,\,0$}} \put(155,121){\mbox{
$V_{\0}$}} \put(216,120){\mbox{$0$}}
 \put(119,0){$0$} \put(159,0){$L$}  \thicklines
\put(120,10){\line(0,1){90}} \put(125,10){\line(0,1){90}}
\put(130,10){\line(0,1){90}} \put(135,10){\line(0,1){90}}
\put(140,10){\line(0,1){90}} \put(145,10){\line(0,1){90}}
\put(150,10){\line(0,1){90}} \put(155,10){\line(0,1){90}}
\put(120,100){\line(1,0){40}} \put(160,100){\line(0,-1){90}}
\put(120,100){\line(1,1){40}} \put(125,100){\line(1,1){40}}
\put(130,100){\line(1,1){18}} \put(160,130){\line(1,1){10}}
\put(135,100){\line(1,1){18}} \put(165,130){\line(1,1){10}}
\put(140,100){\line(1,1){18}} \put(170,130){\line(1,1){10}}
\put(145,100){\line(1,1){18}} \put(175,130){\line(1,1){10}}
\put(150,100){\line(1,1){18}} \put(180,130){\line(1,1){10}}
\put(155,100){\line(1,1){40}}  \put(160,100){\line(1,1){40}}
\put(120,10){\line(1,0){40}}
\put(10,100){\line(1,0){110}}\put(10,10){\line(1,0){110}}
\put(160,100){\line(1,0){110}}\put(160,10){\line(1,0){110}}
\end{picture}
\end{center}

\vspace*{.5cm}


In
general, for arbitrary $\theta$, spin flip contributions occur for
any given incoming polarization.
 Only when $\theta=0$, i.e. for right-angle impingement does spin
flip completely disappear, and this applies both to the step and
barrier  potentials. This limit case reproduces exactly our
previous  one-dimensional results\cite{Ddif}.

Our primary objective in this paper is to present the more general
planar results, i.e. those for arbitrary $\theta$. In doing so, we
observe that spin flip is rigourously absent for the {\em
transmitted wave} in the case of the \emph{barrier potential}.
This is not a resonance phenomena, for even in the so called
``particle'' limit, in which infinite transmitted (and reflected)
waves occur, no spin flip is found. This limit is that pertaining
to an incoming wave packet which is small in $z$ compared  to the
barrier width ($L$). We shall calculate this particle limit by
means of the two step method\cite{And,Sdif} and explicitly verify
the absence of spin flip in transmission. The sum of the infinite
individual waves reproduces the plane wave results (maximum
interference) including resonance phenomena.

Before passing in the next section to a detailed discussion of our
planar diffusion, we wish to recall here that for the
one-dimensional Dirac equation three forms of interaction occur
with a potential, be it a step or barrier of height $V_{\0}$,
depending upon the energy of the incoming particle $E$. For
$E>V_{\0}+m$ diffusion ($\mathcal{D}$), for $V_{\0}-m<E<V_{\0}+m$
tunneling barrier ($\mathcal{T}$), and  for $E<V_{\0}-m$ Klein
energy  zone ($\mathcal{K}$). In the diffusion\cite{Ddif} and
Klein\cite{Kre,Dkle} cases oscillatory solutions exist everywhere.
Whereas the tunneling case\cite{Rec,Dtun} is characterized by real
exponential solutions in the potential region. The interpretation
of the Klein case conventionally involves pair production as an
interpretation of the Klein paradox\cite{Kle}. Only the diffusion
case interests us in this paper although most (but not all)
formulas can be analytically continued into the other energy
zones.

\section*{II. GENERAL CONSIDERATIONS}

The time independent Dirac planar ($y$-$z$) equation for potential
``scattering'' is easily reduced to a one-dimensional problem when
the potential is only a function of one (say $z$) of the planar
variables. Separation of variables results in
\[
\Psi(\boldsymbol{r},t) = \psi(z)\,\exp[i(p_{\2}y-Et)]\,\,,
\]
where $p_{\2}$ is the momentum component along the $y$-direction
and remains so in all regions of the plane. In free space, $V=0$,
$\psi(z)$  satisfies
\begin{equation}
-i\,\alpha_{\3}\,\psi'(z) + \alpha_{\2}\, p_{\2}\,\psi(z) +\beta
\, m\,\psi(z) = E\,\psi(z)\,\,.
\end{equation}
With  $p_{\3}$ the $z$-component of momentum, we have as one of
the explicit solutions (polarized in the z-direction)
\[
\psi(z)=
[\,1,\,\,0,\,\,p_{\3}/(E+m),\,\,ip_{\2}/(E+m)\,]^t\,\,\exp[ip_{\3}z]\,\,
, \] up to an overall normalization factor. We note that
$p_{\2}/p_{\3}=\tan \theta$ and
$E=\sqrt{p_{\2}^{^{2}}+p_{\3}^{^{2}}+m^{\2}}$. Throughout this
paper we consider the incoming particle (travelling from negative
z) polarized as above, but our results will be indifferent to the
specific choice of polarization and indeed will be expressed in
terms of spin conserving and spin flip.

For the region in  which $V(z)=V_{\0}$, we must make the following
modifications (translation)
\[
E  \rightarrow  E-V_{\0}\, \, \,, \,\,\,\,\, p_{\2}  \rightarrow
p_{\2}\, \, \,, \,\,\,\,\, p_{\3}  \rightarrow  q_{\3}\, \, ,
\]
with $(E-V_{\0})^{^{2}}=p_{\2}^{\2}+q_{\3}^{\2}+m^{\2}$. The
existence of a non zero value for $p_{\2}$ modifies our spinors
when compared to our previous one-dimensional calculations, and
hence our diffusion results\cite{Ddif}. It also complicates the
kinematic conditions for being in one of the zones $\mathcal{D}$,
$\mathcal{T}$ and  $\mathcal{K}$. The simplest way to see this is
to define a new mass $m_*$,
\begin{equation}
m_*=\sqrt{p_{\2}^{\2}+m^{\2}}=\sqrt{E^{^{2}}\sin^{\2}\theta +
m^{\2}\cos^{\2} \theta}\,\,.
\end{equation}
Then, the one-dimensional energy zones are simply generalized to
\[
\begin{array}{cclcccl}
(\mathcal{D}) & ~:~~~~~ &  & &  E & > & V_{\0}+m_*\,\,,\\
(\mathcal{T}) & ~:~~~~~ & V_{\0}-m_* & < &  E & < & V_{\0}+m_*\,\,,\\
(\mathcal{K}) & ~:~~~~~ & V_{\0}-m_*  & > &  E & .  &
\end{array}
\]
In Figs.\,(1) and (2), we plot the separation of the various zones
by fixing respectively the incident angle $\theta$ and the
potential $V_{\0}$. The Klein zone is absent for $V_{\0}<2m$, see
Fig.\,(2). The plots in Fig.\,(2) also show that for a given
energy $E$, we may transit for high $E$ from $\mathcal{D}$ to
$\mathcal{T}$ or for low $E$ from $\mathcal{K}$ to $\mathcal{T}$
by varying the incidence angle $\theta$. For
$V_{\0}-m<E<V_{\0}+m$, only the $\mathcal{T}$ zone exists
independent of $\theta$. We note that by varying $\theta$, we can
{\em never} pass through all three zones. We also note that the
one-dimensional kinematics are simply given by the energy points
on the axis $\sin^{\2}\theta=0$ in Fig.\,(2).

The Dirac equation, being of first order in the spatial
derivatives, implies that continuity equations are applied only to
the field. However, this provides \emph{four} equations, one for
each component of the  spinors involved. The ``small'' components
correspond in the non relativistic (NR) limit to the continuity of
the spatial derivative of the NR field. The remaining doubling of
continuity equations (compared to the Schr\"odinger equation) is
exactly what is needed to determine both the spin conserving and
spin flip contributions.

In the next section, we apply the continuity condition to the step
at $z=0$ for waves coming from either direction (needed in the
two-step procedure) and at $z=L$. Below, we sketch the \emph{side
view} for the step calculations of planar diffusion:

\vspace*{.5cm}

\hspace*{-.7cm}
\begin{picture}(480,90) \thinlines
\put(50,10){\vector(0,1){68}} \put(44,80){$V(z)$}
\put(102,25){$V_{\0}$} \put(2,10){\line(1,0){46}}
 \put(0,10){\vector(1,0){103}}
\put(106,8){$z$} \put(48,0){$0$} \put(10,17){\mbox{\small \sc
region I}} \put(56,17){\mbox{\small \sc region II}}
 \thicklines
\put(50,10){\line(0,1){20.5}} \put(50,30){\line(1,0){49}}
 \thinlines
\put(200,10){\vector(0,1){68}} \put(194,80){$V(z)$}
\put(203,25){$V_{\0}$} \put(152,10){\line(1,0){46}}
 \put(150,10){\vector(1,0){103}}
\put(256,8){$z$} \put(197,0){$L$} \put(157,17){\mbox{\small \sc
region II}} \put(206,17){\mbox{\small \sc region III}}
 \thicklines
\put(200,10){\line(0,1){20.5}} \put(200,30){\line(-1,0){49}}
 \thinlines
\put(350,10){\vector(0,1){68}} \put(344,80){$V(z)$}
\put(402,25){$V_{\0}$} \put(302,10){\line(1,0){46}}
 \put(300,10){\vector(1,0){103}}
\put(406,8){$z$} \put(348,0){$0$} \put(310,17){\mbox{\small \sc
region I}} \put(356,17){\mbox{\small \sc region II}}
 \thicklines
\put(350,10){\line(0,1){20.5}} \put(350,30){\line(1,0){49}}
\thicklines \put(18,61){\vector(1,0){30}}
\put(52,50){\vector(1,0){30}}
 \put(48,50){\vector(-1,0){30}}
\put(52,30){\vector(1,0){30}}
 \put(48,39){\vector(-1,0){30}}
\put(12,58){$1$}
\put(-9,47){$R_{\pl}(0)$}\put(85,47){$T_{\pl}(0)$}
\put(-9,36){$R'_{\pl}(0)$}\put(85,36){$T'_{\pl}(0)$}
\put(168,61){\vector(1,0){30}} \put(202,50){\vector(1,0){30}}
 \put(198,50){\vector(-1,0){30}}
\put(162,58){$1$}
\put(139,47){$R_{\mi}(L)$}\put(235,47){$T_{\mi}(L)$}
\put(202,39){\vector(1,0){30}}
 \put(198,39){\vector(-1,0){30}}
\put(139,36){$R'_{\mi}(L)$}\put(235,36){$T'_{\mi}(L)$}
\put(381.5,61){\vector(-1,0){30}} \put(352,50){\vector(1,0){30}}
 \put(348,50){\vector(-1,0){30}}
\put(385,58){$1$}
\put(293,47){$\widetilde{T}_{\mi}(0)$}\put(385,47){$\widetilde{R}_{\mi}(0)$}
\put(352,39){\vector(1,0){30}}
 \put(348,39){\vector(-1,0){30}}
\put(293,36){$\widetilde{T}'_{\mi}(0)$}\put(385,36){$\widetilde{R}'_{\mi}(0)$}
\put(80,80){\mbox{\small \sc \fbox{step a}}}
\put(230,80){\mbox{\small \sc \fbox{step b}}}
\put(380,80){\mbox{\small \sc \fbox{step c}}}
\end{picture}

\vspace*{.5cm} \noindent
 In these figures, we list the contributing reflection ($R$) and
 transmission ($T$) amplitudes. The $\pm$ subscript distinguish an
 up step from a down step while the tilde represents reflection and transmission
 for an incoming wave from the  \emph{right}. The prime terms correspond
 to spin flip amplitudes. In the next section, we also give the
 results of the continuity equations for the barrier.

\section*{III. PLANE WAVE RESULTS}

Consider the step potential defined by
\[ V(z) =
\left\{\,0\,\,\,\mbox{for}\,\,z<0\,\,,\,\,\,V_{\0}\,\,\,
\mbox{for}\,\,z>0\,\right\}\,\,,
\]
with $V_{\0}>0$ and with an incoming plane wave from the left (see
the step A) with a definite polarization (spin along the
$z$-direction). The spinor continuity equations are

\begin{eqnarray}
 \left(\,\begin{array}{c}E+m\\0\\
p_{\3}\\i\,p_{\2}
\end{array}\,\right)\,
 \exp[\,ip_{\3}z\,] + \left[\, R_{+}(0)\, \left(\,\begin{array}{c} E+m\\ 0\\
 - p_{\3}\\ i p_{\2}
\end{array}\,\right) +
R'_+(0)\, \left(\,\begin{array}{c} 0\\ E+m\\
 - ip_{\2}\\ p_{\3}
\end{array}\,\right)\,\right]\, \exp[\,-\,ip_{\3}z\,]  \nonumber \\
= \frac{E+m}{E-V_{\0}+m}\,\,
\left[\,T_+(0)\,\left(\,\begin{array}{c}E-V_{\0}+m\\0\\
q_{\3}\\i\,q_{\2}
\end{array}\,\right) +
T'_+(0)\,\left(\,\begin{array}{c} 0 \\E-V_{\0}+m\\
-i\,q_{\2} \\ -q_{\3}
\end{array}\,\right)
\,\right]\,
 \exp[\,iq_{\3}z\,]\,\,.
\end{eqnarray}

\noindent These matrix equations can be rewritten as
\[
\left[
\begin{array}{c}
(E+m)\,[1+R_{+}(0)]\\
(E+m)\,R'_{+}(0)\\
p_{\3}[1-R_+(0)]-ip_{\2} R'_{+}(0)\\
ip_{\2}[1+R_+(0)]+p_{\3} R'_{+}(0)
\end{array}
\right] = \,\frac{E+m}{E-V_{\0}+m}\,\,\left[
\begin{array}{c}
(E-V_{\0}+m)\,T_+(0)\\
(E-V_{\0}+m)\,T'_{+}(0)\\
q_{\3}T_+(0)-iq_{\2}T'_{+}(0)\\
iq_{\2}T_+(0)-q_{\3}T'_{+}(0)
\end{array}
\right]
\]
Solution of which yields
\begin{eqnarray}
R_+(0) & = &
\frac{(p_{\2}^{\2}+m^{\2}+mE)\,V_{\0}}{(E+m)(p_{\3}^{\2}+p_{\3}q_{\3}-V_{\0}E)}\,
\, ,
\nonumber \\
T_+(0) & = & \frac{p_{\3}^{\2}(E-V_{\0}+m)+p_{\3}q_{\3}(E+m)}{
(E+m)(p_{\3}^{\2}+p_{\3}q_{\3}-V_{\0}E)} \, \, ,\nonumber \\
R'_+(0) & = &i\,
\frac{p_{\2}p_{\3}\,V_{\0}}{(E+m)(p_{\3}^{\2}+p_{\3}q_{\3}-V_{\0}E)} \, \, ,\nonumber \\
T'_+(0) & = & R'_+(0)\, \, .
\end{eqnarray}
Here, we observe that because the step is situated at $z=0$, the
$R_+(0)$ and $T_+(0)$  amplitudes are real while the $R'_+(0)$ and
$T'_+(0)$ are imaginary. All four amplitudes exist. If we change
the incoming polarization (from spin up to spin down), we find
exactly the same solutions, although from a different form of
matrix equation. This is a general property and we shall
henceforth not repeat this observation.

As a simple check of our results, we note that when $V_{\0}\to 0$
only  $T_+(0)=1$ survives as must be. We also observe that as
$p_{\2}\to 0$ ($\theta \to 0$)  the spin flip terms vanish and
this confirms our one-dimensional results. The other two step
cases (B and C) can be calculated in the same way and yield the
results
\begin{eqnarray}
R_-(L) & = & -\, \frac{[p_{\2}^{\2}+m^{\2}+m(E-V_{\0})]\,V_{\0}}{
(E-V_{\0}+m)(p_{\3}^{\2}+p_{\3}q_{\3}-V_{\0}E)}\,\exp[2iq_{\3}L] \,\, ,\nonumber \\
T_-(L) & = & \frac{q_{\3}^{\2}(E+m)+p_{\3}q_{\3}(E-V_{\0}+m)}{
(E-V_{\0}+m)(p_{\3}^{\2}+p_{\3}q_{\3}-V_{\0}E)}\,\exp[i(q_{\3}-p_{\3})L]
 \,\, ,\nonumber \\
 R'_-(L) & = &-\,i\,
\frac{p_{\2}q_{\3}\,V_{\0}}{(E-V_{\0}+m)(p_{\3}^{\2}+p_{\3}q_{\3}-V_{\0}E)}\,
\exp[2iq_{\3}L] \,\, ,  \nonumber \\
 T'_-(L) & = &
 R'_-(L)\,\exp[-\,i(q_{\3}+p_{\3})L] =
 R'_-(0)\,\exp[i(q_{\3}-p_{\3})L]\,\,.
\end{eqnarray}
While for right impingement at $z=0$ we obtain,
\begin{equation}
\widetilde{R}_-(0) = R_-(0)\,\,\,,\,\,\,\,\,  \widetilde{T}_-(0)
=T_-(0) \, \, \, , \, \, \, \, \,  \widetilde{R}'_-(0)
=-R'_-(0)\,\,\, ,\,\,\,\,\, \widetilde{T}'_-(0) =
\widetilde{R}'_-(0)\,\,.
\end{equation}

Passing now to the barrier, we define the reflection and
transmission amplitudes respectively in regions I and III by $R$,
$R'$, $T$ and $T'$, while for region II, that of the potential
$V_{\0}$, we use $A$, $A'$, $B$ and $B'$. These are not related in
a simple manner to our previous step results. The connection will
however be derived in the next section.

The expression for $\psi(z)$ within each region are given below.
\mbox{\small \sc Region I} $(z < 0)$:
\begin{equation}
 \left(\,\begin{array}{c}E+m\\0\\
p_{\3}\\i\,p_{\2}
\end{array}\,\right)\,
 \exp[\,ip_{\3}z\,] + \left\{\, R\, \left(\,\begin{array}{c} E+m\\ 0\\
 - p_{\3}\\ i p_{\2}
\end{array}\,\right) +
 R'\, \left(\,\begin{array}{c} 0\\ E+m\\
 - ip_{\2}\\ p_{\3}
\end{array}\,\right)\,\right\}\, \exp[\,-\,ip_{\3}z\,]\,\,.
\end{equation}
\mbox{\small \sc Region II} $(0<z <L)$:
\begin{eqnarray}
\left\{\,A\,\left(\,\begin{array}{c}E-V_{\0}+m\\0\\
q_{\3}\\i\,q_{\2}
\end{array}\,\right) +
A'\,\left(\,\begin{array}{c} 0 \\E-V_{\0}+m\\
-i\,q_{\2} \\ -q_{\3}
\end{array}\,\right)
\,\right\}\,
 \exp[\,iq_{\3}z\,]\,\, + \nonumber \\
 \left\{\,B\,\left(\,\begin{array}{c} E-V_{\0}+m\\ 0 \\
-q_{\3}\\i\,q_{\2}
\end{array}\,\right) +
B'\,\left(\,\begin{array}{c} 0 \\E-V_{\0}+m\\
-i\,q_{\2} \\ q_{\3}
\end{array}\,\right)
\,\right\}\,
 \exp[\,-iq_{\3}z\,]\,\,.
\end{eqnarray}
\mbox{\small \sc Region III} $(z > L)$:
\begin{equation}
\left\{\,T\,\left(\,\begin{array}{c}E+m\\0\\
p_{\3}\\i\,p_{\2}
\end{array}\,\right)+
 T'\, \left(\,\begin{array}{c} 0\\ E+m\\
 - ip_{\2}\\ -p_{\3}
\end{array}\,\right)\,\right\}\, \exp[\,ip_{\3}z\,]\,\,.
\end{equation}
In the above expressions factors such as $1/(E+m)$ and
$1/(E-V_{\0}+m)$ have been absorbed into the amplitudes for
simplification. After elimination of the intermediate $A$, $A'$,
$B$ and $B'$ the coupled continuity equations yield the matrix
equation
\[
\left[
\begin{array}{c}
(E+m)\,(1+R)\\
(E+m)\,R'\\
p_{\3}(1-R)-ip_{\2}R'\\
ip_{\2}(1+R)+p_{\3}R'
\end{array}
\right] = M \,\left[
\begin{array}{c}
(E+m)\,T\\\
(E+m)\,T'\\
p_{\3}T-ip_{\2}T'\\
ip_{\2}T-p_{\3}T'
\end{array}
\right] \,\exp[\,ip_{\3}L] \,\,,
\]
where
\[ M =
 \frac{1}{q_{\3}}\, \left[
\begin{array}{cccc}
q_{\3}\cos(q_{\3}L) & q_{\2} \sin(q_{\3}L) & - i (E-V_{\0}+m)
\sin(q_{\3} L) & 0\\
q_{\2} \sin(q_{\3}L) & q_{\3}\cos(q_{\3}L) & 0 & i (E-V_{\0}+m)
\sin(q_{\3} L)\\
- i (E-V_{\0}-m) \sin(q_{\3} L) & 0 & q_{\3}\cos(q_{\3}L) & q_{\2}
\sin(q_{\3}L) \\
0 & i (E-V_{\0}-m) \sin(q_{\3} L) & q_{\2} \sin(q_{\3}L) &
q_{\3}\cos(q_{\3}L)
\end{array}
\right]\,\, .
\]
The solution of these equations are
\begin{eqnarray}
R & = & - \,i \,
\left(m+\frac{p_{\2}^{^{2}}}{E+m}\right)V_{\0}
\, \frac{\sin(q_{\3}L)}{p_{\3}q_{\3}\mathcal{F}} \,\,,\nonumber \\
R' & = & \frac{p_{\2}p_{\3}}{E+m}\, V_{\0}\,
\frac{\sin(q_{\3}L)}{p_{\3}q_{\3}\mathcal{F}}\,\,, \nonumber \\
T & = & \frac{\exp(-ip_{\3}L)}{\mathcal{F}}\,\,, \nonumber\\
T'& = & 0\,\,,
\end{eqnarray}
with
\[ \mathcal{F} = \cos(q_{\3}L) - \,i \,
\frac{p_{\3}^{^{2}} -EV_{\0}}{p_{\3}q_{\3}}\,\sin(q_{\3}L)\,\,.
\]
These are the generalized barrier results for a plane wave. They
contain the momentum $p_{\2}$ indicating a dependence upon
incident angle. For $p_{\2}=0$, we reproduce the one-dimensional
Dirac barrier results published in our previous paper\cite{Ddif}.
Spin flip is indeed absent in this limit. However, the surprising
feature of the general barrier results is that $T'=0$ for all
incident angles. This was not expected and seems to contrast with
the fact that the equivalent terms $T'_{\pm}$ are not identically
null for the steps.

In the next section, we perform the two step calculation to
redetermine the above expressions, in particular that for $T'$.
This will confirm the above results and demonstrate that $T'=0$ is
{\em not} a resonance phenomena. On the other hand typical
resonance behavior is present in the above expressions. Whenever
$\sin(q_{\3}L)=0$, i.e. for $q_{\3}L=n \pi$ (with $n$ a positive
integer) both reflected amplitudes, $R$ and $R'$, vanish and the
transmitted probability $|T|^{{^2}}=1$. Plane waves are
theoretical abstractions, the above results are in truth good
approximations only for {\em barrier widths much smaller than the
incoming  wave packets widths}.  In other words high values of $n$
corresponding to large $L$ will {\em not} exhibit resonance
behavior\cite{Ddif}.

\section*{IV. TWO STEP CALCULATION}

In this section, we recalculate the barrier results using the step
results. Specifically, this approach for the barrier is called the
two step method and does not \emph{directly} involve the
continuity equations of the previous section.  The method uses
three step results. In addition to that for the step at $z=L$
impinged upon from the left, it also uses the step results at
$z=0$ twice. Once for the initial incoming wave (impingement from
the left) and then those for a wave reflected from the end of the
barrier with consequent impingement at $z=0$ from the right.

This method of calculation can be used to reproduce the standard
barrier results by simply adding the infinite contributions to
transmission and reflection yielding the so called (plane) wave
limit.  By treating each contribution as incoherent with the
others we obtain the particle limit. Probabilities are conserved
in both limits although the total transmission/reflection
probabilities are quite different. The wave limit for example is
characterized by resonance phenomena, the particle limit is not.
In addition to calculating the individual particle limit
probabilities for transmission, it is one of our objectives to
control if vanishing spin flip is a resonance phenomena or not.

The incoming wave is polarized. At $z=0$ it encounters the first
of the two steps and \emph{two} contributions proceed (are
transmitted) to the right. These are indicated by the spin
conserving $T_+(0)$ and the spin flip term $T'_+(0)$. At the
second (downward) step at $z=L$ \emph{each } of these
contributions produces two transmitted amplitudes. These four
transmitted terms combine into two sums: the spin conserving
amplitude,
\begin{equation}
\label{at} A_{\T}  = T_+(0)T_-(L)+ T'_+(0)T'_-(L)  \,\, ,
\end{equation}
and a spin flip amplitude,
\begin{equation}
A'_{\T}=T_+(0)T'_-(L) + T'_+(0)T_-(L)\,\,.
\end{equation}
Now, we can use the step results previously given to observe that
\begin{equation}
A_{\T} =
\frac{2\,p_{\3}q_{\3}}{p_{\3}^{\2}+p_{\3}q_{\3}-E\,V_{\0}}\,
\exp[i\,(q_{\3}-p_{\3})\,L]\,\,\,\,\,\,\,\mbox{and}\,\,\,\,\,\,\,
A'_{\T} =  0\,\,.
\end{equation}
So the first particle spin flip contribution is {\em null}. This
fact alone does not guarantee that subsequent spin flip
contributions are null. For example, the second transmitted
contributions contain in addition to the $T$ factors also two $R$
factors corresponding to an additional back and forth passage over
the barrier,
\begin{eqnarray}
A_{\R}  & = & R_-(L)\widetilde{R}_-(0)+ R'_-(L)\widetilde{R}'_-(0)
= \frac{(E^{^{2}}-p_{\3}^{\2})\,V_{\0}^{^{2}}
}{(p_{\3}^{\2}+p_{\3}q_{\3}-E\,V_{\0})^{^{2}}}
\,\exp[2\,i\,q_{\3}\,L]\,\,,\\
A'_{\R}  & = & R_-(L)\widetilde{R}'_-(0)+
R'_-(L)\widetilde{R}_-(0) =  0\,\,.
\end{eqnarray}
The overall second contribution to the spin flip is thus
\[ A_{\T}A'_{\R}+A'_{\T}A_{\R}=0\,\,.\]
All higher spin flip contributions take the form of the the above
second contribution multiplied by powers of the spin-conserving
double reflection factor $A_{\R}$. Thus, the vanishing of the
second  spin flip term does indeed imply the vanishing of
\emph{all} the spin flip contributions.

We now calculate the individual spin conserving contributions. The
first contribution $A_{\T}$ has already been given (\ref{at}). The
second contribution is
\[
A_{\T}A_{\R}+A'_{\T}A'_{\R}=A_{\T}A_{\R}\,\,.
\]
All higher (later emerging) contributions are now obvious. The
$n$-th term reads,
\[A_{\T}A^{^{n-1}}_{\R}\,\,.\]
In the \emph{wave limit} these contributions must be added
coherently to give a single outgoing transmission amplitude. This
may conveniently be written as,
\begin{equation}
 \frac{A_{\T}}{1-A_{\R}}\,\,.
\end{equation}
After inserting the specific step expressions of the previous
section, we reproduce after a little algebra the plane wave result
for the barrier transmission,
\begin{equation}
T=
\frac{A_{\T}}{1-A_{\R}}=\frac{\exp(-ip_{\3}L)}{\mathcal{F}}\,\,.
\end{equation}

\section*{V. CONCLUSIONS}

In the literature one normally encounters one dimensional
potential analysis. However, when spin and relativity are relevant
one-dimensional analysis may be too limited. For example the
absence of spin flip terms for one-dimensional step and barrier
diffusion is no longer valid for planar diffusion where an angle
of incidence exists.  The potentials are still considered
functions of a single spatial variable ($z$ in this paper) but the
incoming particle have two momentum components. This small
modification produces significant differences, specifically the
appearance, in general, of spin flip terms and consequent
modifications in the non-flip amplitudes. We have demonstrated
these facts in this paper. These effects are a direct consequence
of the angular dependence of the Dirac spinors.

We have found one notable exception to the above, for which we
have no \emph{simple} explanation, although we believe one must
surely exist. This exception is that in the case of the barrier
potential there is \emph{no spin-flip transmission amplitude}.
There are always spin-flip terms for the step, be the step rising
or dropping , except of course for the one dimensional limit when
the potential is met head on. This makes the barrier result all
the more unexpected, since we know that the barrier result may be
derived from a double-step analysis which uses only the step
results.

We have listed in the previous sections all relevant planar
diffusion results for the step and barrier. We have also derived
the transmission amplitude for the barrier in the "particle" limit
via the two-step method. This demonstrates that the absence of
spin-flip for \emph{each} individual outgoing wave packet,
independent of the degree of coherence involved, i.e. it is
\emph{not} a resonance type phenomena.

As a side product, we have described the kinematics of dirac
planar scattering and observed that by varying the incidence
angle, for a given incoming energy we may transit through two
kinematic zones, e.g. from diffusion to or from tunnelling, but
never through all three kinematic zones $\mathcal{D}$,
$\mathcal{T}$, $\mathcal{K}$. We have in the past referred to
these zones as \emph{energy zones}, but this is correct only for
one-dimensional studies. They should instead be referred to as
kinematic zones which depend upon both energy and angle of
incidence. Energy alone, normally does not determine these zones.
These kinematic zones are distinguished by quite distinct physics.
In this paper, we have limited our attention to diffusion.


\newpage

\begin{figure}[hbp]
\hspace*{-2.5cm}
\includegraphics[width=19cm, height=24cm, angle=0]{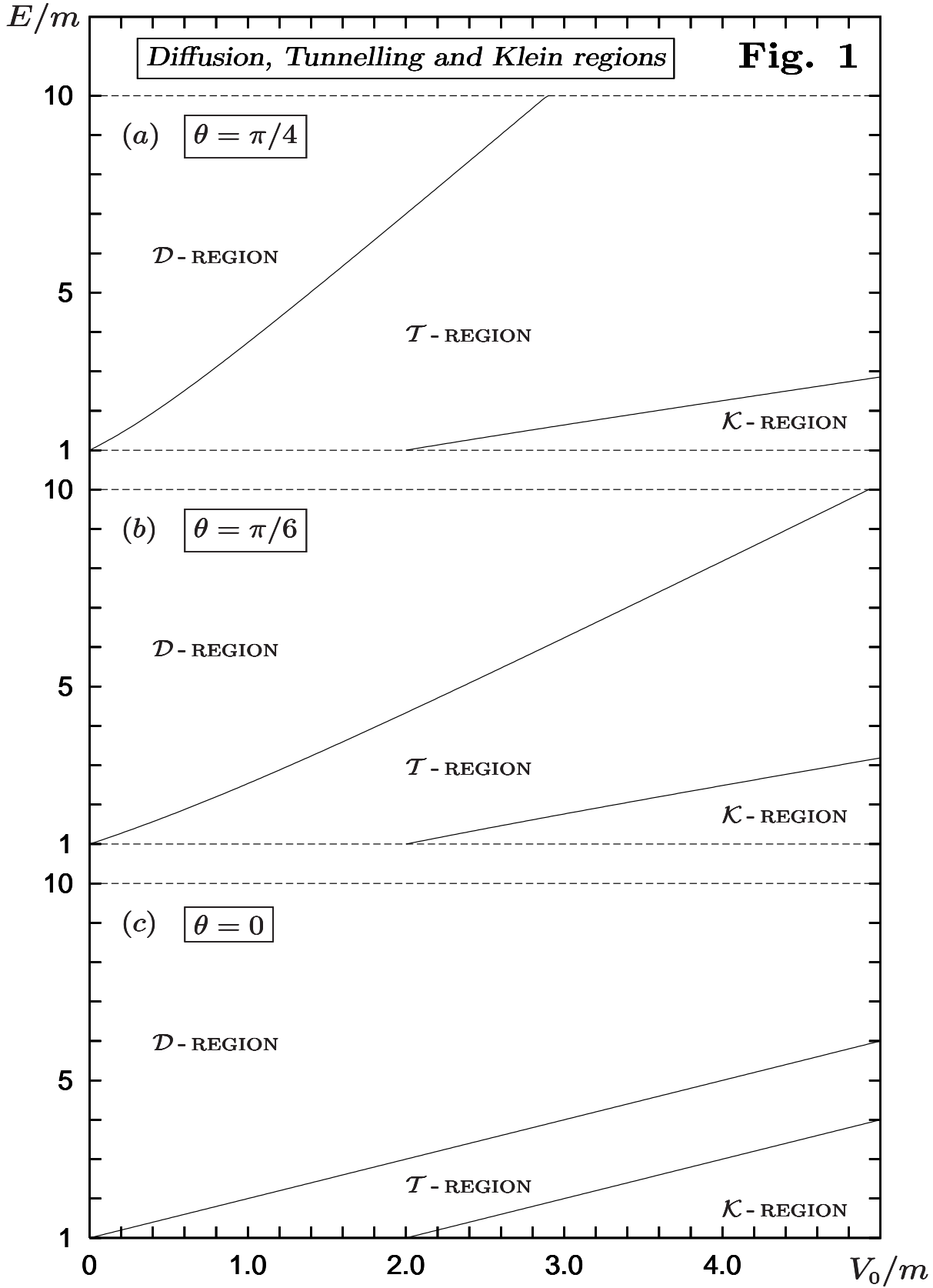}
\vspace*{-1.8cm}
 \caption{The separate kinematic zones for three choices of incoming
 angle as a function of potential height. Case (c) corresponds to the one-dimensional plot.}
\end{figure}

\newpage

\begin{figure}[hbp]
\hspace*{-2.5cm}
\includegraphics[width=19cm, height=24cm, angle=0]{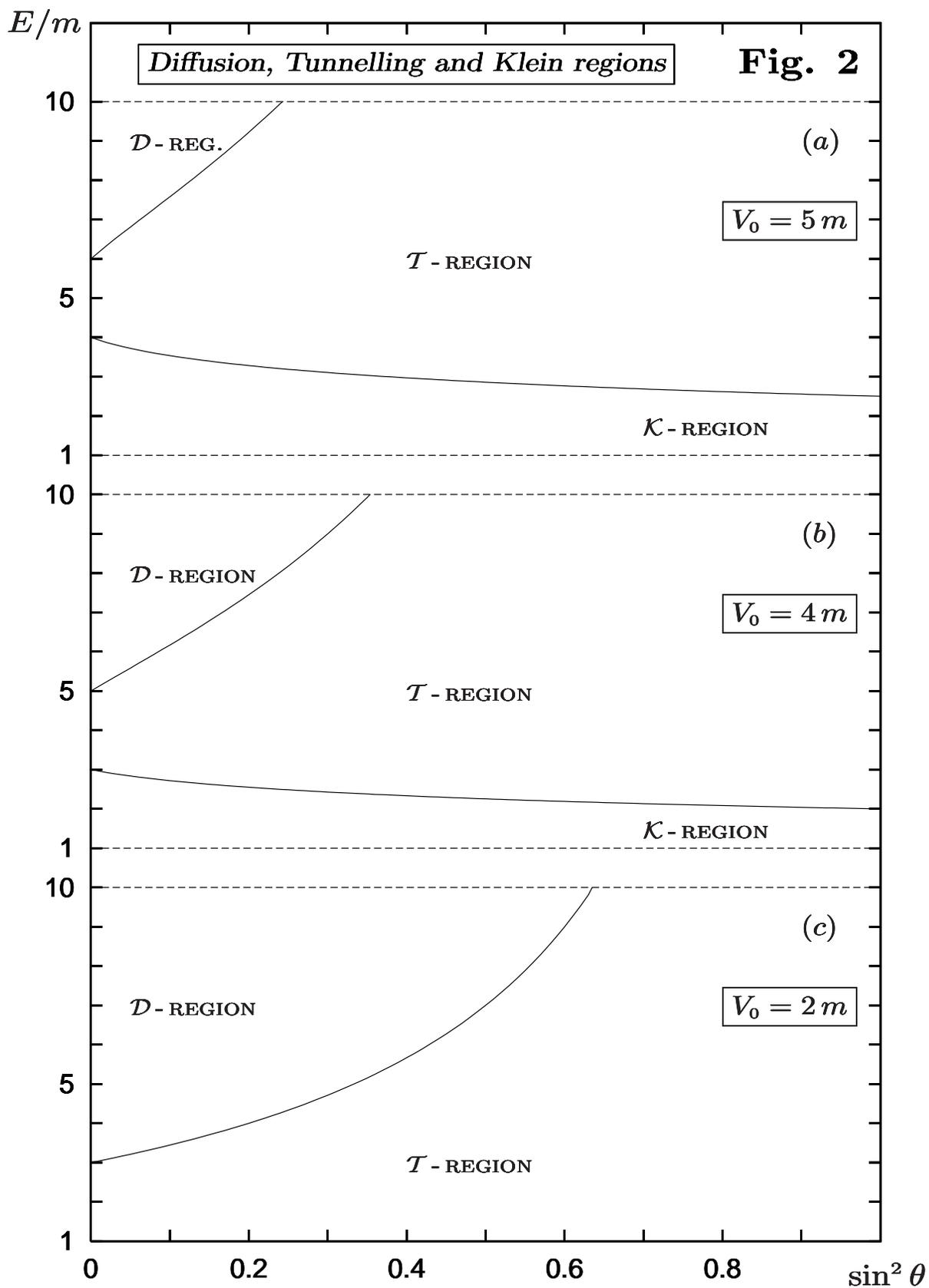}
\vspace*{-1.8cm}
 \caption{The angular dependence separated into kinematic zones for three choices of potential.
           Case (c) is the limit case (highest potential) for which there is no Klein zone.}
\end{figure}

\end{document}